\documentclass[mypaper,7pt,twoside]{CoAst}
\usepackage{graphicx,fancyhdr,floatflt}
\renewcommand{\chaptermark}[1]{\markboth{#1}{}}

\newcommand{\kopf}{\small\itshape Comm. in Asteroseismology\\ Vol. 148, 2006}
\newcommand{\Authors}[1]{\begin{center}\normalsize\bf\sf #1 \end{center}}

\renewcommand{\author}[1]{\begin{center}\normalsize\bf\sf #1 \end{center}}
\newcommand{\Address}[1]{\begin{center}\small\sf #1 \end{center}}

\newcommand{\References}[1]{\begin{flushleft}{\large References\\}\vspace*{2mm}\small #1 \end{flushleft}}

\newcommand{\chapterDSSN}[2]{\chapter[\sf\normalsize #1\\ \footnotesize \hspace*{5mm}by #2 \sf\normalsize][]{#1\\}\rhead[\fancyplain{}{\sf\footnotesize \center{#1}}]{\fancyplain{}{\sffamily\thepage}}\lhead[\fancyplain{\kopf}{\sffamily\thepage}]{\fancyplain{\kopf}{\sf\footnotesize \center{#2}}}}

\renewcommand{\chaptername}{}
\newcommand{\acknowledgments}[1]{\vspace*{5mm}\noindent\begin{bf}Acknowledgments. \end{bf} #1}

\begin{document}
\sf

\chapterDSSN{Amplitude Saturation in $\beta$ Cephei Models - Preliminary Results}{R. Smolec, P. Moskalik}

\Authors{R. Smolec$^1$, P. Moskalik$^1$} 
\Address{$^1$ Copernicus Astronomical Centre, Bartycka 18, 00-716 Warsaw, Poland}

\noindent

We present preliminary results concerning amplitude saturation in $\beta$~Cephei models. Using nonlinear approach we have investigated amplitude limitation mechanism in $\beta$~Cephei stars. In our approach radial modes have been treated as representative for all acoustic oscillations. We have studied pulsation properties of several models (7--20\thinspace M$_\odot$, $Z=0.02,\ 0.015$) using radiative Lagrangean hydrocodes (essentially those of Stellingwerf 1974, 1975). Nonlinear limit cycles (monoperiodic full-amplitude oscillations) have been calculated through Stel\-lin\-gwerf's (1974) relaxation technique, which also provides information about limit cycle stability. 

In our main survey ($Z=0.02$) only the fundamental and the first overtone modes are linearly excited. Nonlinear growth rates have been used to determine the modal selection (see {\it e.g.} Stellingwerf 1975). We found that fundamental mode pulsation is dominant. First overtone pulsation is restricted to intermediate masses and to the vicinity of the blue edge. The first overtone and the fundamental mode pulsation domains are separated by either-or or narrow double-mode domains.

Predicted, single mode saturation amplitudes have been compared to amplitudes observed for monoperiodic $\beta$~Cephei variables (Fig.~1, left). Predicted amplitudes are significantly higher. The amplitudes may be lowered by decreasing the metal abundance of the models, $Z$. We have found that for $Z=0.015$, decrease of model amplitudes is not sufficient. At the same time instability strip shrinks and leaves a lot of stars beyond the blue edge. By lowering $Z$ we are not able to match simultaneously the observed amplitudes and the instability strip. 

Predicted amplitudes may be easily lowered to the observed level if one assumes collective saturation of the pulsation instability, by $n$ similar acoustic modes. In this hypothetical multimode solution, amplitudes of individual modes are a factor of $\sim\sqrt{n}$ lower than in single mode solution. Using linear code of Dziembowski (1977) we have determined the number of linearly unstable acoustic modes for models of different masses, located in the center of the main sequence band. This number doesn't vary much along evolutionary track and thus, was assumed to be representative for all models of a given mass. The number of unstable modes is much higher than the number of detected modes in the multiperiodic $\beta$~Cephei variables. Nonlinear simulations (Nowakowski 2005) also show that not all unstable modes take part in the saturation process. Thus, we have arbitrarily assumed that only one third of linearly unstable modes take part in the saturation. Amplitudes rescaled under assumption of collective saturation are presented in Fig.~1 (right). Using only part of the linearly unstable acoustic modes, we have lowered theoretical amplitudes to the observed level. Thus, we argue that collective instability saturation is sufficient to explain observed amplitudes of the $\beta$~Cephei pulsators. A possible difficulty of this model is that predicted pulsation-induced broadening of spectral lines might be higher than observed. We discuss this problem in Smolec \& Moskalik (2006).
\begin{figure}[t]
\includegraphics[width=11.8cm]{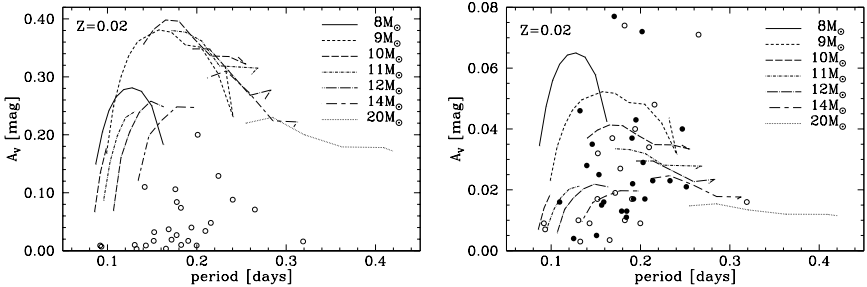}
\caption{Predicted single mode saturation amplitudes (left) and amplitudes assuming collective saturation by several acoustic modes (right). For comparison, amplitudes of monoperiodic (open circles) and multiperiodic (filled circles) $\beta$~Cephei stars are ploted.}
\end{figure}

In several of our radiative models we have found numerically robust, double-mode behaviour, with radial fundamental and first overtone modes simultaneously excited. This form of pulsation is encountered only in intermediate mass models (10--11\thinspace M$_\odot$). Depending on the specific model, the origin of double mode pulsation can be traced to one of two different mechanisms: either to the nonresonant coupling of the two excited modes, or to the $2\omega_1 \simeq \omega_0 + \omega_2$ parametric resonance.

Full results of this analysis (including discussion of non-uniform filling of the theoretical instability strip by $\beta$~Cephei variables and detailed study of the double-mode models) are presented in Smolec \& Moskalik (2006).

\acknowledgments{The authors are grateful to the EC for the establishment of the European Helio- and Asteroseismology Network HELAS, which made the participation of the authors at this workshop possible. This work has been supported by the Polish MNiI Grant No. 1 P03D 011 30.}

\References{
Dziembowski, W.A. 1977, AcA 27,95\\
Nowakowski, R. 2005, PhD Thesis, Copernicus Astronomical Center, Warsaw\\
Smolec, R., Moskalik, P. 2006, MNRAS, submitted\\
Stellingwerf, R.F. 1974, ApJ 192, 139\\
Stellingwerf, R.F. 1975, ApJ 195, 441\\

}

\end{document}